# On Simplified 3D Finite Element Simulations of Three-core Armored Power Cables

**Juan Carlos del-Pino-López [1,\*], Marius Hatlo [2] and Pedro Cruz-Romero [1]**

[1] Electrical Engineering Department, Escuela Técnica Superior de Ingeniería, Universidad de Sevilla, Camino de los Descubrimientos s/n, 41092, Sevilla, Spain; vaisat@us.es; plcruz@us.es
[2] Technological Analysis Center, Nexans Norway AS, Halden, Norway; marius.hatlo@nexans.com
\* Correspondence: vaisat@us.es; Tel.: +34-954-552-847

**Abstract:** This paper analyzes different ways to simulate electromagnetically three-core armored cables in 3D by means of the finite element method. Full periodic models, as lengthy as 36 m, are developed to evaluate the accuracy when simulating only a small portion of the cable, as commonly employed in the literature. The adequate length and boundary conditions for having the same accuracy of full periodic models are also studied. To this aim, five medium voltage and high voltage armored cables are analyzed, obtaining the minimum length of the cable that may be simulated for having accurate results in shorter time and with less computational burden. This also results in the proposal of a new method comprising the advantages of short geometries and the applicability of periodic boundary conditions. Its accuracy is compared with experimental measurements and the IEC standard for 145 kV and 245 kV cables. The results show a very good agreement between simulations and measurements (errors below 4 %), obtaining a reduction in the computation time of about 90 %. This new method brings a more effective tool for saving time and computational resources in cable design and the development of new analytical expressions for improving the IEC standard.

**Keywords:** three-core; armor; 3D; finite element method; cable twisting.

## 1. Introduction

In the last decades there has been an important development in offshore wind farms due to higher and steadier mean wind speeds [1,2]. This growth is helping in improving the technology and efficiency of all the components involved, hence reducing the cost of these expensive installations. In this sense, one of the main components that have an important impact in the total cost are the submarine power cables needed for exporting the power energy. Its size, and consequently its cost, is defined by the IEC 60287 standard [3]. However, for the particular case of three-core armored cables, many studies have concluded that they are usually oversized, since the IEC standard overestimates the power losses in these cables [4-9]. The main reason of these results is because the analytical expressions provided in the IEC standard date back to the 1940's [10], and later updated in the 1990's [11], where semi-empirical equations were derived by considering the armor as a tube enclosing the three phases (like pipe-type cables). This geometry allows currents to circulate along the "tube" armor, so armor losses may be of importance. However, this assumption ignores the fact that the armor is composed by steel wires which are helically twisted over the three phases. This way the relative position of active and passive conductors changes along the cable length. Thus, the net induced voltage in each armor wire is zero due to the twisting [4]. Consequently, no net current may flow along the armor wires in balanced systems as assumed erroneously in the IEC standard, being the power losses in the armor only due to eddy currents within each individual wire and hysteresis losses.

Nonetheless, although researchers agree in the need of updating the expressions of the IEC standard, [9,12] report some contradictory results regarding the importance of screen (sheath) over armor losses in the global computation of cable losses. In this sense, [6-7] conclude that armored cables have higher losses than unarmored cables. However, the additional losses do not occur only in the armor itself, but also in the three screens. The main reason of this conclusion is that the presence of the armor disturbs the magnetic field distribution inside the cable, which is "compressed" under the armor, leading to higher flux linking the screens. Therefore, larger circulating currents are induced, and higher losses result in the screens [13]. This effect may be also influenced by the non-linear properties of the armor wires, which is something still to be analyzed in depth. Therefore, it may be of interest not only to update the IEC expression of the armor's loss factor ($\lambda_2$), but also that of the screen's loss factor ($\lambda_1$), as suggested in [9,13,14].

To this aim, accurate tools are required to simulate and analyze the complex physics and interactions involved in three-core armored cables, since experimental measurements are costly and only possible for manufacturers. In this sense, numerical simulations based on the finite element method (FEM) have been employed widely in the past [4, 6, 9, 12-16]. This tool is usually applied to 2D geometries, sometimes representing the armor as a hollow cylinder with appropriate electrical properties to include the presence of gaps between the armor wires [15-18]. In any case, 2D FEM simulations do not consider the helical twisting of cores and armor wires, since 2D geometries assume that all cores and armor wires are straight and parallel. To include the twisting of passive and active conductors, [4] proposed one of the most extended solutions, known as 2.5D FEM simulations, where the armor wires are connected in series through an external electric circuit which is coupled to the FEM model. This ensures a total armor current to be zero. Alternatively, some studies have proposed new analytical approaches for the estimation of power losses in three-core armored cables, such as [16]. Also, other studies propose improvements based on the method of moments (MoM-SO) [18-19] and the use of sub-conductor equivalent circuits [20], but they can only handle a constant relative permeability for the armor wires. Nonetheless, both numerical and analytical tools are commonly based on 2D geometries, where only the effect of the transverse magnetic field is considered. Subsequently, these approaches do not include the magnetic field component that is parallel to the armor wires [19]. This is of particular importance since it is the main cause of the eddy currents induced in the magnetic armor, and it is clearly influenced by the relative twisting between cores and armor (usually twisted with different lay lengths and in the opposite direction to achieve torsion stability). Hence, the mentioned approaches may provide wrong induced voltages along the screens and armor wires, resulting in false induced currents circulating within them.

Therefore, for a precise modeling of three-core armored cables, a 3D analysis would be required to accurately compute the fields produced by the helical paths of active and passive conductors in the cable. However, few studies have employed 3D geometries of three-core armored cables since simulations are very time consuming and require powerful computers. In this sense, some studies have employed several simplifications on 3D FEM models. For example, in [9] a basic 3D FEM model was developed in COMSOL Multiphysics© [21] with 10 straight armor wires around three twisted conductors assumed as edges, where the wires were assumed as non-conductive to simplify the complexity of the geometry and reduce the size of the mesh. Alternatively, in [22] a 1200 mm$^2$ three-core submarine cable with 119 armor wires and lead screens was modelled using the same software. Non-linear magnetic properties were employed in the armor wires. The model length was only about 1/3 of the core's lay length to reduce the computational cost. The results derived from this 3D model confirm the higher losses provided by the IEC standard. However, it is not clear if the length of this model is acceptable to compute the exact steady-state solution of actual cables. In this sense, [23] employs a 3-m long 3D FEM model to check both IEC standard and 2.5D FEM simulations. Although the results obtained from this shortened model seem to be in good agreement with experimental measurements [24], this work also states that the model length is not completely adequate to represent the phenomena that takes place in this cable, since a 6-m long geometry would be required for having a model length that provides periodicity in both the phases and armor lay length (the least common multiple (*LCM*) of both lay lengths). However, due to the high demanding computational requirements, authors claim in [20,23,24] that 3D models larger than 3 m cannot be solved by any FEM software. But in [25] a 3D FEM model as long as 14 m was developed in COMSOL Multiphysics© for obtaining the series resistance and the inductive reactance of a 800 mm$^2$ 145 kV three-core submarine power cable. The model length was equal to the *LCM* of cores and armor lay length, and periodic boundary conditions were applied at both ends of the geometry, obtaining results in good agreement with experimental measurements. [25] also presented the feasibility of 3D FEM models to develop a complete parametric analysis by showing the influence of the armor permeability and the core and armor twisting on different aspects such as power losses and self, mutual and sequence impedances.

From all these studies it is clearly concluded that 3D FEM models provide very important data and valuable knowledge about all phenomena that take place inside three-core armored cables, being an essential tool for cable design and for the development of new accurate analytical expressions of the current rating in these power cables. However, after reviewing the state of the art, one question arises in relation to the simulation of these cables: is full periodicity certainly required in 3D FEM models? Indeed, the use of periodic models as long as the *LCM* of phases and armor lay length may not always help in reducing the size of the model (i.e. a cable with a lay length of 3.3 for the cores and 4 m for the armor would require a 132 m model length). Therefore, in many situations extremely large computational resources would be required to solve the model.

To put some order in this matter, this work analyzes in Section 2 the accuracy of non-periodic models when compared to full periodic models, also looking for the shortest length to be considered for having accurate results. In Section 3 a new short 3D periodic model is proposed to reduce the size and complexity of 3D FEM models, so that lower computational resources may be required for having the same accuracy as full periodic models. The error in power losses as well as the reduction in simulation times are also compared to those provided by non-periodic models. Eventually, Section 4 compares this new proposal with experimental measurements and the IEC standard.

## 2. Accuracy of non-periodic 3D FEM models

For three-core armored cables, the model size can be reduced by taking advantage of symmetries in the geometry, such as axial periodicity, so that only a short portion of the cable may be simulated for having the same results as infinitely long models. In this sense, an appropriate axial length for the model (named henceforth periodic length) should be selected to ensure periodicity in both the phases and armor lay length (an integer number of turns for both the armor wires and the phases). Thus, its relative position at both ends of the geometry remains equal, and both ends are also equally oriented, as represented in Figure 1 by two coordinate systems ($\vec{e}_x$, $\vec{e}_y$, $\vec{e}_z$) and ($\vec{e}'_x$, $\vec{e}'_y$, $\vec{e}'_z$). Therefore, the destination boundary is a simple geometric translation of the source boundary along the *z* axis, so it is easy for the FEM software to apply periodicity by just forcing the solution of the magnetic vector potential *A* at each destination point ($A_{dst}$) to be equal to the solution at a corresponding source point ($A_{src}$).

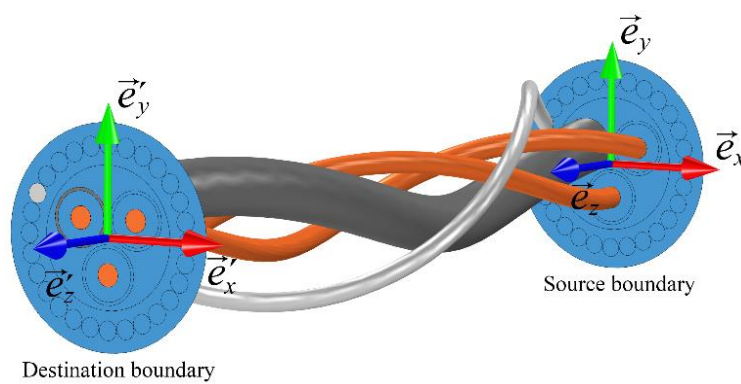

**Figure 1.** Boundary orientation in periodic geometry.

However, when the length selected for a 3D FEM model leads to a non-periodic model, the relative position between phases and armor wires is different at both ends of the geometry, thus no periodic boundary conditions can be applied there. This forces the induced currents in the armor to flow differently than they would do in a periodic geometry, especially at both ends, since Ampere's Law must be fulfilled. This results in a different distribution of the power losses in the armor, as shown in Figure 2, where they are represented for the cases of periodic boundary conditions (Figure 2a) or non-periodic (Figure 2b). As can be observed, while Figure 2a shows a complete periodicity in the distribution of the losses, Figure 2b presents a different behavior at each end of the cable. This is quantified in Figure 3 where the relative difference between the armor losses obtained in both situations is represented in %. It is to be noticed that, in the periodic case, there are regions at both ends of the geometry where the armor losses may be more than 80 % higher than those of the non-periodic case.

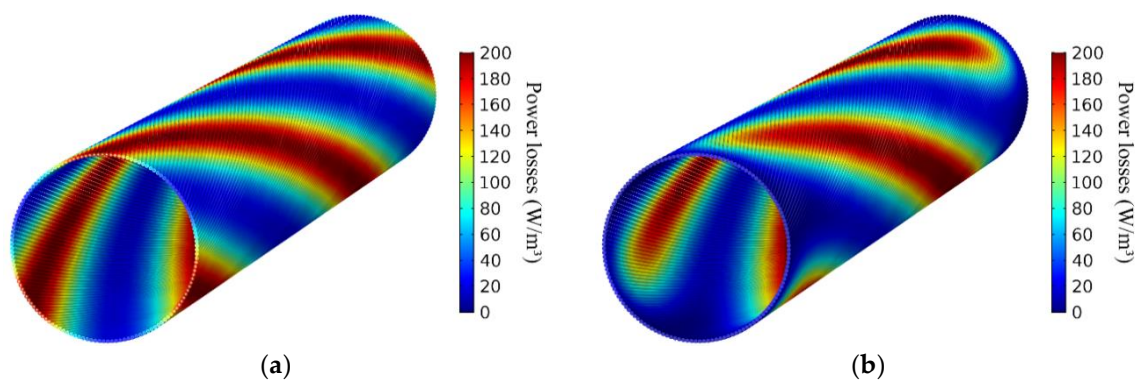

**Figure 2.** Distribution of armor losses in: (**a**) periodic FEM model; (**b**) non-periodic 3D FEM model.

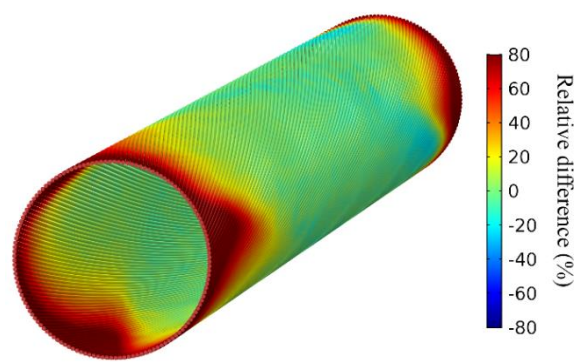

**Figure 3.** Relative difference in the armor losses between periodic and non-periodic 3D FEM models.

Therefore, the influence of these regions in the computation of the armor losses may be of importance, depending on their relative size to the periodic length of the model, which also depends on the relative twisting of conductors and armor wires and the selected length for the non-periodic model. In other words, smaller errors are to be expected in larger models, since these regions would have less relative weight in the total armor losses. However, this may be computationally inefficient, so it is of interest to find the minimum length that may be employed in non-periodic 3D FEM models that ensures accurate results. To clarify this point, a set of 3D FEM simulations have been carried out for five medium voltage (MV) and high voltage (HV) three-core armored cables (Table 1). The values of electrical resistivity of conductors and sheaths are taken from the IEC standard [3] and a complex relative permeability for the armor of $\mu_r$ =100 − 50$j$ was used. All models were solved in Comsol Multiphysics©. For each cable, the loss factors for the sheaths ($\lambda_1$) and the armor ($\lambda_2$) have been computed for the five cables by means of full periodic and non-periodic 3D FEM models. The procedure is as follows: $\lambda_1$ and $\lambda_2$ are obtained for different lengths (*L*) of the portion of the cable simulated in non-periodic 3D FEM models (no periodic boundary conditions were applied at both ends of the geometry). The same is done by means of full periodic models, with a periodic length equal to the *LCM* of armor and phases lay length (from 1.2 m to 36 m-length models, as shown in Table 1), so that periodic boundary conditions were applied at both ends of the geometry. Then, taking the results from the full periodic models as the "exact value", the error in $\lambda_1$ and $\lambda_2$ is obtained.

Table 1. MV and HV cable data.

|  | Cable 1 | Cable 2 | Cable 3 | Cable 4 | Cable 5 |
|---|---|---|---|---|---|
| Voltage (kV) | 30 | 145 | 170 | 245 | 245 |
| Conductor | Cu | Cu | Cu | Al | Cu |
| Cross-section (mm$^2$) | 35 | 800 | 630 | 1200 | 1600 |
| Conductor radius (mm) | 3.5 | 17.5 | 15.25 | 21.45 | 23.15 |
| Sheath thickness (mm) | 0.9 | 3.7 | 2.4 | 2.25 | 2.25 |
| Sheath radius (mm) | 9.25 | 43.8 | 39.75 | 49.75 | 52 |
| Core lay length (m) | 1.2 | 2.8 | 3.6 | 4 | 4 |
| Wires diameter (mm) | 6 | 5.6 | 5.6 | 5 | 5.6 |
| Nº of wires | 28 | 114 | 103 | 139 | 129 |
| Armor radius (mm) | 29.32 | 104.5 | 91.91 | 115.6 | 120.75 |
| Armor lay length (m) | 0.4 | 3.5 | 2.2 | 3.6 | 3.6 |
| *LCM* (m) | 1.2 | 14 | 7.2 | 36 | 36 |
| *CP* (m) | 1.2 | 1.56 | 1.37 | 1.89 | 1.89 |

For this task, a workstation composed by two Intel® Xeon E5-2630 2.6 GHz CPUs with 256 Gb of RAM memory and a 2 Tb hard disk drive for memory swapping was employed. The results are represented in Figure 3 as a function of the model length. Since these cables are very different in size, the length of the model is characterized in Figure 4 by means of a more generic parameter that represents the relative twisting between phases and armor. It is the so-called relative pitch length or crossing pitch (*CP*), defined in [7] for armor and phases twisted in opposite directions (contralay) as

$$CP = \frac{1}{\frac{1}{p_a} + \frac{1}{p_c}}, \qquad (1)$$

where $p_a$ and $p_c$ denote the lay length of armor and phases in meters, respectively. Thus, the ratio *L/CP* represents the length employed in the model relative to the *CP* of each cable (Table 1).

From Figure 4a it can be derived that differences below 5 % in relation to full periodic models can be obtained in $\lambda_1$, provided that the length employed for the non-periodic model is, at least, a 75 % of the *CP* of each cable. Moreover, models as short as 50 % of the *CP* may be employed if an error below 10 % can be assumed. Nonetheless, the error in $\lambda_2$ presents a more erratic behavior (Figure 4b). In this case, differences below 10 % are guaranteed by non-periodic models when using a model length larger than 110 % of the *CP*. In fact, for a model length in the order of 125 % of the *CP* a relative error below 5 % may be achieved in most of the cables considered. For this length, Table 2 shows the relative error in the series resistance (*R*) and inductive reactance (*X*) of the cable, the sheath current ($I_s$), the loss factors and the simulation time reduction ($\Delta T$) relative to the full periodic models.

From these results it is easily concluded that short geometries can be employed in non-periodic 3D FEM models for having reliable and meaningful results not only in $\lambda_1$ and $\lambda_2$, but also in other parameters, hence greatly reducing the computational resources and time for the simulations. Nonetheless, a further analysis may provide new ways to obtain more accurate results, comprising the advantages of short geometries and the applicability of periodic boundary conditions, as developed next.

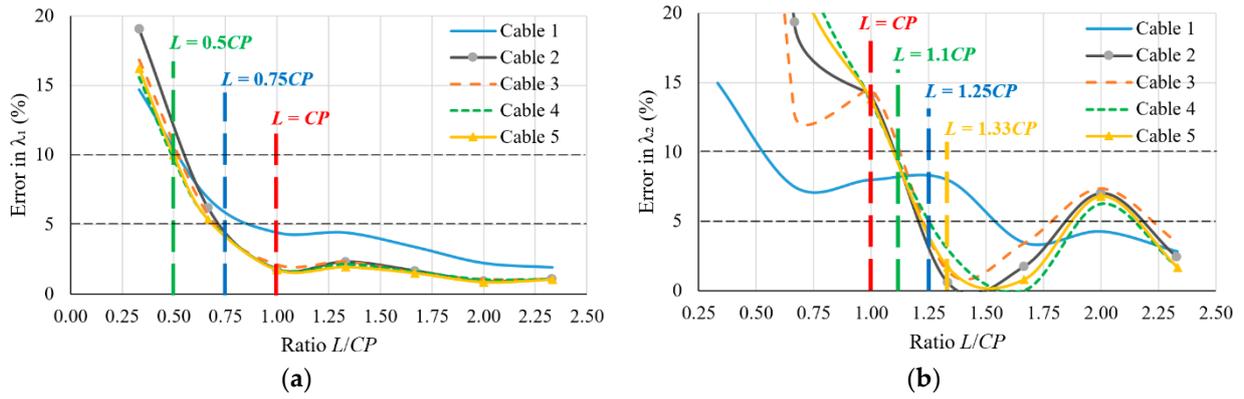

**Figure 4.** Error as a function of the ratio *L/CP* in: (**a**) $\lambda_1$; (**b**) $\lambda_2$.

**Table 2**. Error in *R*, *X*, $I_s$, $\lambda_1$, $\lambda_2$ and simulation time reduction in relation to full periodic models for *L* = 1.25*CP*.

|  | Cable 1 | Cable 2 | Cable 3 | Cable 4 | Cable 5 |
|---|---|---|---|---|---|
| $\varepsilon_R$ (%) | 2.08 | 1.19 | 1.2 | 1.11 | 1.07 |
| $\varepsilon_X$ (%) | 0.8 | 0.98 | 0.8 | 0.69 | 0.88 |
| $\varepsilon_{Isheath}$ (%) | 1.12 | 0.48 | 0.47 | 0.49 | 0.44 |
| $\varepsilon_{\lambda 1}$ (%) | 4.43 | 2.32 | 2.31 | 2.15 | 1.93 |
| $\varepsilon_{\lambda 2}$ (%) | 8 | 2.61 | 2.73 | 5.01 | 4.83 |
| $\Delta T$ (%) | 82 | 96 | 92 | 98 | 98 |

## 3. Proposal of a new 3D short periodic model

As observed previously in Figure 2, the distribution of the power losses in the armor shows a helical and periodic pattern, with higher losses in areas close to the power conductors. This pattern is a consequence of the relative twisting of armor wires and phase conductors, i.e. the *CP* defined earlier. Since this parameter defines the distance where an armor wire meets twice a particular phase conductor, it represents some kind of periodicity in the geometry of the cable. This is observed in Figure 5 for Cable 1, where the magnitude of the magnetic field is represented in different slices that are separated a distance equal to the *CP* (0.3 m). As can be observed, all slices are equal but rotated a certain angle following the phase twisting.

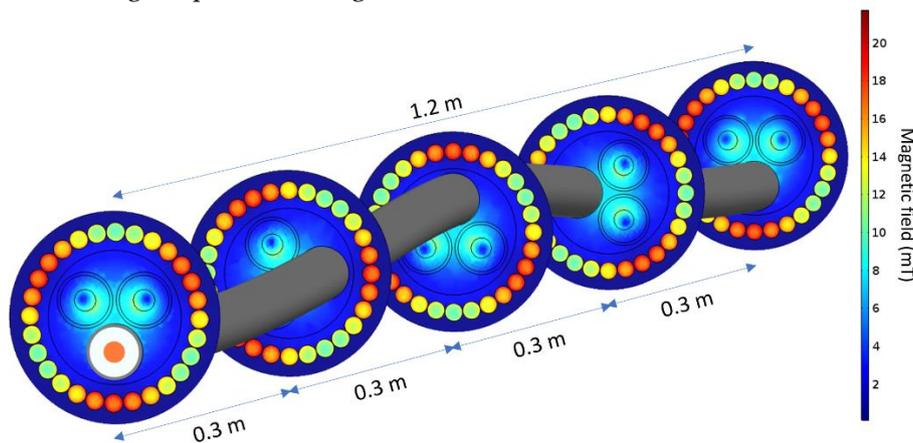

**Figure 5.** Magnetic field magnitude in different slices spaced a distance equal to the *CP* of Cable 1.

Similar conclusions are derived from Figure 6, where the magnetic flux lines trapped inside an armor wire are depicted. It shows how the maximum in the magnetic field inside the wire is periodically reached when the wire is closer to a phase conductor. It also shows that the flux lines enter and leave the wire at certain areas depending on its relative position to the phase conductors, as shown in Figure 7a with more detail. Furthermore, this pattern is repeated every 0.3 m but rotated.

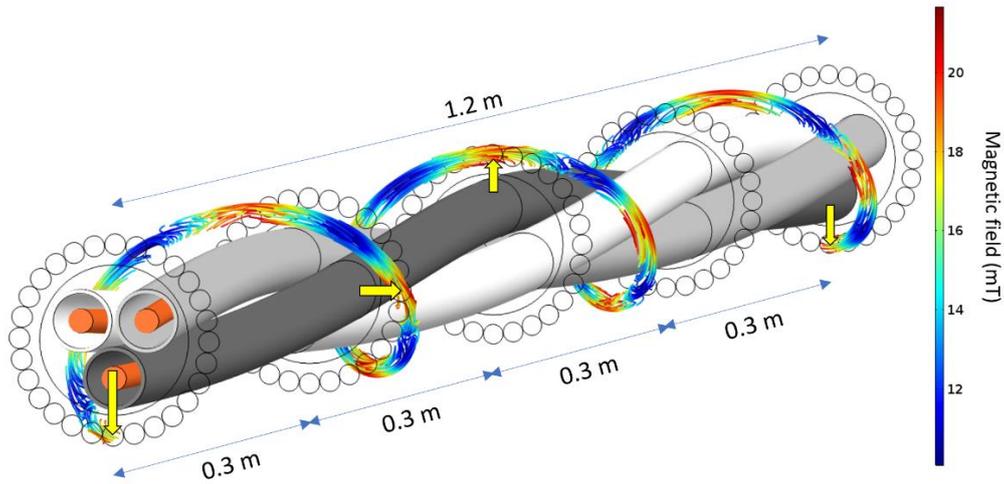

**Figure 6.** Magnetic flux lines along an armor wire in Cable 1.

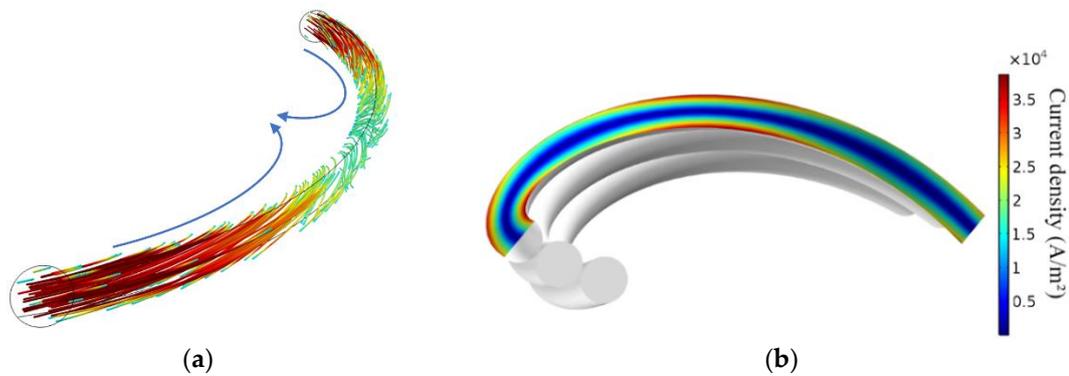

**Figure 7.** Detailed armor wire in Cable 1: (**a**) Magnetic field flow inside the wire; (**b**) Current density distribution in a cross-section of a portion of the wire.

All this must be reflected in the current density distribution inside the armor wires. Certainly, the magnitude of the current density inside the wires is not uniform, as represented in Figure 7b for a cross-section along a portion of a wire, being quite low in the center all along the wire and reaching maximums periodically at certain areas close to the surface. This is repeated along the whole length of the wire, giving rise to the pattern shown in Figure 8 (note: dimensions not to scale), where there are 12 regions where the maximum current density is achieved.

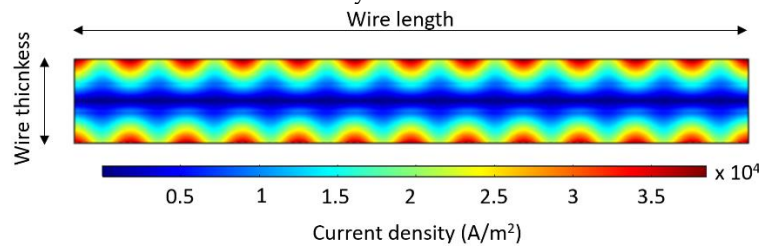

**Figure 8.** Current density distribution along a cross-section of an armor wire in Cable 1 (dimensions not to scale).

This pattern is a consequence of how the induced eddy currents flow inside the armor wire and how its magnitude and flowing direction evolve along the wire length depending on its relative position to the phase conductors. In this sense, Figure 9a shows how eddy currents flow in close paths inside the armor wires and close to its external surface (skin effect), so that the longitudinal net current in the armor is virtually zero. However, they may flow clockwise or counterclockwise at different locations inside the armor wire, as observed in Figure 9b, where the arrows represent the direction of the flow, and its size and color represent the magnitude of the current density inside a cross-section along an armor wire.

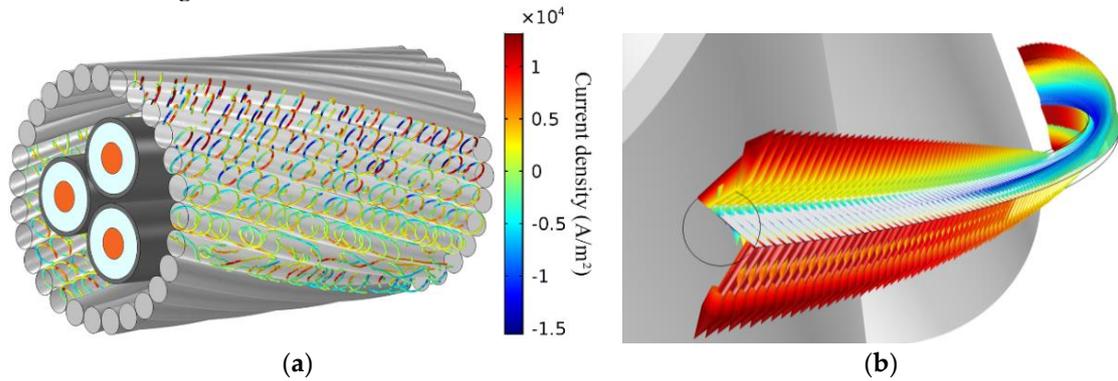

**Figure 9.** (**a**) Eddy currents circulating inside the armor wires in Cable 1; (**b**) Direction of the flowing path of eddy currents inside a portion of armor wire in Cable 1.

As can be seen, the magnitude of the current density is maximum at the beginning of this wire, and currents flow clockwise. Then, it starts decreasing along the wire until a point where it increases again, but now flowing counterclockwise. Furthermore, this pattern is observed on every armor wire and it is repeated every 0.3 m (the *CP*) along the model length but rotated. All this gives rise to the 12 maximums in the current density observed earlier in Figure 8 (each armor wire passes close to the phase conductors three times every 0.3 m, hence repeated four times for the periodic length of 1.2 m).

In summary, the solution of the whole cable can be derived by only analyzing a short portion as lengthy as the *CP* (Figure 10a). Nonetheless, although the relative position between the phase conductors and the armor wires is identical at both ends of this short model, now the source and destination boundaries are not equally oriented, as shown in Figure 10b, so periodic boundary conditions must be properly applied. Since in three-core armored cables the armor and the phase conductors are usually twisted in different directions and with a different lay length, it is hard for the FEM software to find the transformation matrix that encodes the relative orientation of the source and destination points to apply periodicity. It is then needed to set up manually its relative orientation by assigning an appropriate coordinate system for each boundary.

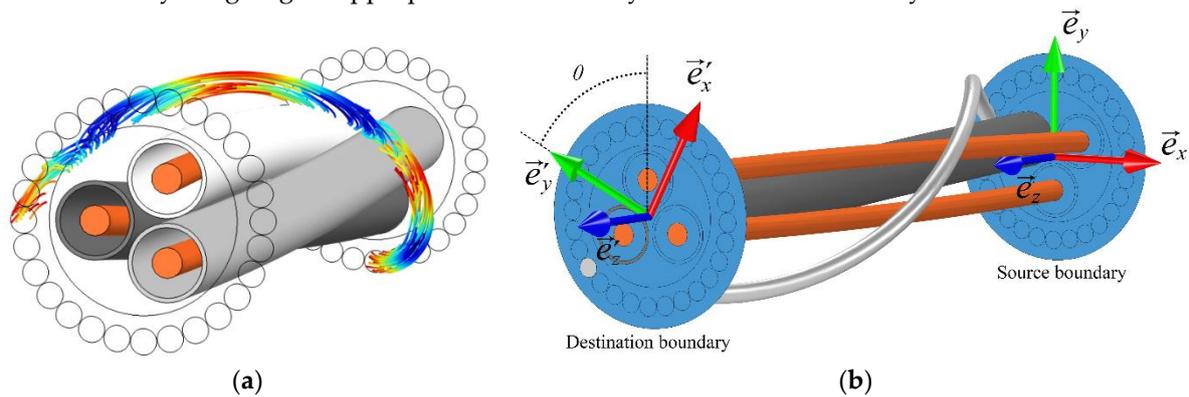

**Figure 10.** (**a**) Short periodic model for Cable 1; (**b**) Orientation of boundaries in short periodic model for Cable 1.

To this aim, the global coordinate system ($\vec{e}_x$, $\vec{e}_y$, $\vec{e}_z$) is assigned to the source boundaries and a new one ($\vec{e}'_x$, $\vec{e}'_y$, $\vec{e}'_z$) is obtained for the destination boundaries by rotating the global system a certain angle $\theta$ (Figure 10b):

$$\vec{e}'_x = \cos\theta \cdot \vec{e}_x + \sin\theta \cdot \vec{e}_y,$$
$$\vec{e}'_y = -\sin\theta \cdot \vec{e}_x + \cos\theta \cdot \vec{e}_y, \quad (2)$$
$$\vec{e}'_z = \vec{e}_z,$$

being

$$\theta = \pm 2\pi \frac{CP}{p_c}, \qquad (3)$$

where the plus sign denotes phases twisted in counterclockwise and the minus sign for the clockwise case.

*3.1. Comparison with full periodic and non-periodic models*

This new short periodic model has the advantage of having a length equal to the *CP*, less that the length of 1.25*CP* suggested earlier for the case of non-periodic models. From this point of view this new model would take shorter time to simulate. However, the use of rotated periodic boundary conditions increases the complexity of the model and hence the solution time. To clarify whether short periodic or non-periodic model is faster and more accurate, Table 3 presents some results, including the error in *R*, *X*, $I_s$, $\lambda_1$, $\lambda_2$, as well as the reduction in simulation time when compared to the full periodic model ($\Delta T_f$) and the non-periodic model ($\Delta T_{np}$) of each cable.

**Table 3**. Error in *R*, *X*, $I_s$, $\lambda_1$, $\lambda_2$ and simulation time reduction in relation to full periodic and non-periodic models with *L* = 1.25*CP*.

|  | Cable 1 | Cable 2 | Cable 3 | Cable 4 | Cable 5 |
|---|---|---|---|---|---|
| $\varepsilon_R$ (%) | 0.15 | 0.08 | 0.02 | 0.01 | 0 |
| $\varepsilon_X$ (%) | 0 | 0.07 | 0.07 | 0.09 | 0.11 |
| $\varepsilon_{Isheath}$ (%) | 0.9 | 0.03 | 0.02 | 0.07 | 0.05 |
| $\varepsilon_{\lambda 1}$ (%) | 0.34 | 0.08 | 0.01 | 0.05 | 0.04 |
| $\varepsilon_{\lambda 2}$ (%) | 0.07 | 0.18 | 0.09 | 0.08 | 0.08 |
| $\Delta T_f$ (%) | 83 | 97 | 93 | 99 | 99 |
| $\Delta T_{np}$ (%) | -7.6 | 17 | -14 | 0.9 | 4.5 |

From these results it should be remarked the small error in the electrical parameters and the loss factors obtained with the new short periodic model when compared to the full periodic one (most of them below 0.5 %), especially if one has in mind the great differences in size, complexity and simulation time between both models. This proves that this short periodic model behaves "exactly" like the full model, providing lower error than those obtained by non-periodic models. Moreover, when compared to 1.25*CP* length non-periodic models, Table 3 shows that, although there are no important differences in simulation time, the new short periodic model is often faster than non-periodic models (negative value in *ΔT<sub>np</sub>* denotes faster computation in non-periodic models).

*3.2. Computational requirements*

The great reduction in simulation time brings the opportunity of using less expensive equipment for the simulations, and also the possibility of considering additional aspects to make the model more accurate. For example, the use of a complex permeability for the armor wires ($\mu_r$) as a function of the magnetic field *f(B)* as defined in [9]

$$\mu_r = \mu_{0r} + \mu_{mr}\left(1 - e^{-\alpha_1 |B|}\right) - j\mu_{mi}\left(1 - e^{-\alpha_2 |B|}\right), \qquad (4)$$

To illustrate this, some results are presented in Table IV for Cable 2 (Figure 11a) and Cable 4 (Figure 11b) when using a full 3D periodic model (F3D) and the new short 3D periodic model here proposed (S3D). Also, some results obtained from 2.5D FEM simulations are included. The size of the model is indicated by its length, as well as the number of mesh elements employed and the number of degrees of freedom to be solved (DoFs). The simulation time and the memory requirements are also included for two different computers: a workstation composed by two Intel® Xeon E5-2630 2.6 GHz CPUs with 256 Gb of RAM memory, and an Intel® i7-4770K 3.5 GHz CPU with 32 Gb of RAM memory. Both computers have a 2 Tb hard disk drive for memory swapping. The parameters for Equation (4) are selected from [9] for grade 34.

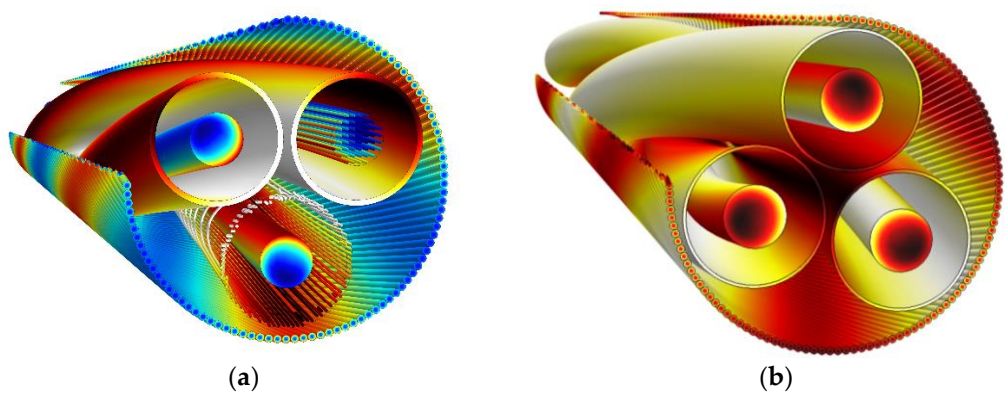

(**a**)  (**b**)

**Figure 11.** Current density distribution in short 3D periodic models for: (**a**) Cable 2; (**b**) Cable 4.

As expected, 3D models take longer time than few seconds, as 2.5D models do. However, it can be observed the great reduction obtained in computing time and memory requirements when using the approach proposed in this work (S3D). In particular, it can be observed how the 14-m long 3D FEM model of Cable 2 toked about 30 hours to be solved when using an armor permeability of $\mu_r$ = $f(B)$, being necessary to use the hard disk drive for memory swapping. Conversely, with the proposed approach (S3D) it toked only 1.5 hours and using just a 23 % of the available RAM memory resources. Furthermore, if a constant permeability of $\mu_r$ = 100 − 50$j$ is employed for the armor, then time reduces to just 15 minutes for the same mesh size. Table 4 also shows that the F3D model of Cable 4 toked about 17 hours to be solved, but it reduces to less than 1 hour in the workstation when using S3D for a finer mesh than that employed in the F3D model. On the other hand, Table 4 also shows that both cables can be now simulated in a smaller computer when using S3D, hence reducing the need of large computational resources for analyzing such cables with good accuracy.

**Table 4**. Model size and solution time.

| | Size | Length (m) | Nº Mesh elements | DoFs | $\mu_r$ | CPU | Time | Memory (Gb) |
|---|---|---|---|---|---|---|---|---|
| **Cable 2** | F3D | 14 | 3·10$^6$ | 3.2·10$^6$ | $f(B)$ | E5 | 30 h | 256+swap |
| | S3D | 1.55 | 6.5·10$^5$ | 1.2·10$^6$ | $f(B)$ | E5 | 1.5 h | 60 |
| | | | | | 100-50$j$ | E5 | 15 min | 60 |
| | | | | | 100-50$j$ | i7 | 30 min | 32+swap |
| | 2.5D | - | 9.2·10$^4$ | 4.6·10$^4$ | $f(B)$ | i7 | 8 s | 1 |
| **Cable 4** | F3D | 36 | 8.3·10$^6$ | 18·10$^6$ | 100-50$j$ | E5 | 17 h | 256+swap |
| | S3D | 1.89 | 9.5·10$^6$ | 1.8·10$^6$ | 100-50$j$ | E5 | 50 min | 90 |
| | | | | | 100-50$j$ | i7 | 76 min | 32+swap |
| | 2.5D | - | 9·10$^4$ | 4.5·10$^4$ | 100-50$j$ | i7 | 7 s | 1 |

**4. Comparison with experimental measurements and the IEC standard**

In the following, simulations developed by S3D are compared to experimental measurements for cables 2, 4 and 5. The electrical and magnetic properties of the cables were obtained from experimental measurements and adjusted for the correct temperature [6], [26]. For Cable 4 two different materials are considered for the armor: galvanized steel (with magnetic properties) and stainless steel (with no magnetic properties). For the three cables the per unit length resistance R and the inductive reactance X are obtained, comparing the results with experimental measurements in two situations: armor and screens solidly bonded and in open circuit. All the results and the relative differences ($\varepsilon_R$ and $\varepsilon_X$) between S3D simulations and measurements are presented in Table 5 and Figure 12, where results obtained from 2D and 2.5D FEM models are also included as a reference. It is to be remarked that, even though there are always uncertainties related to measurements and simulations, the results here presented are in very good agreement for all the situations analyzed, with relative errors always lower than 4 %. These results show the accuracy and the usefulness of the short 3D periodic model here proposed, both in *R* and *X*, especially when compared to the results provided by 2D and 2.5D FEM model, as shown in Figure 12a, where 2.5D models present the greatest error in the resistance and 2D models in the reactance.

**Table 5**. Experimental and S3D simulation results.

| | Bonding | Result | R (Ω/km) | $\varepsilon_R$ (%) | X (Ω/km) | $\varepsilon_X$ (%) |
|---|---|---|---|---|---|---|
| **Cable 2** | Solid | Measured | 0.0455 | 0.22 | 0.12 | 1.66 |
| | | FEM | 0.0454 | | 0.118 | |
| **Cable 4[1]** | Solid | Measured | 0.042 | 0.47 | 0.114 | -0.17 |
| | | FEM | 0.0418 | | 0.1142 | |
| | Open | Measured | 0.0332 | -0.30 | 0.118 | -0.34 |
| | | FEM | 0.0333 | | 0.1184 | |
| **Cable 4[2]** | Solid | Measured | 0.038 | 2.63 | x | x |
| | | FEM | 0.037 | | 0.11 | |
| | Open | Measured | 0.031 | 3.55 | x | x |
| | | FEM | 0.0299 | | 0.112 | |
| **Cable 5[1]** | Solid | Measured | 0.0314 | -1.27 | 0.107 | -0.94 |
| | | FEM | 0.0318 | | 0.108 | |
| | Open | Measured | 0.0221 | -3.17 | 0.111 | -1.35 |
| | | FEM | 0.0228 | | 0.1125 | |

[1] Galvanized steel; [2] Stainless steel.

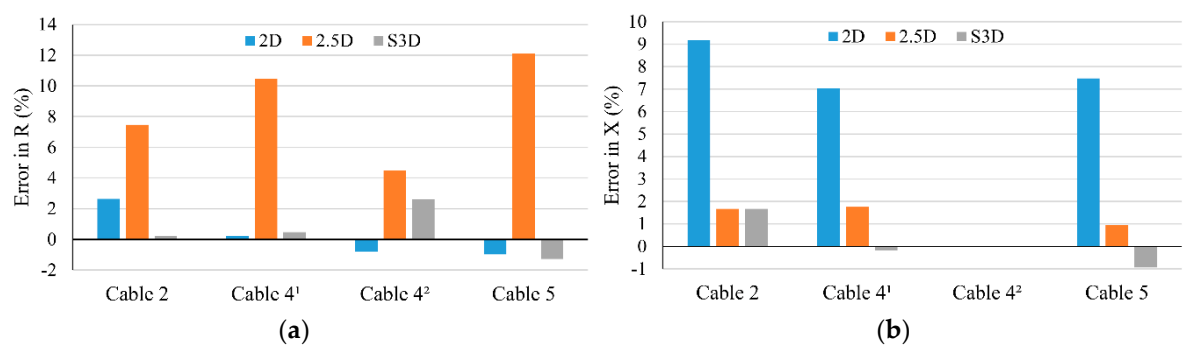

**Figure 12.** Errors in different FEM models relative to experimental measurements (solid bonding) for: (**a**) *R*; (**b**) *X*.

Regarding the power losses, Figure 13 presents the results for the screens and armor loss factors ($\lambda_1$ and $\lambda_2$) derived not only from different FEM models, but also from the IEC standard and estimates resulting from experimental measurements (screens and armor solidly bonded). In the latter case, $\lambda_1$ and $\lambda_2$ are obtained only for cables 4 and 5 by means of the procedure presented in [5,7], that requires measurements in the cable with and without the armor [26]. It should be remarked that this procedure only provides estimates of $\lambda_1$ and $\lambda_2$, since it is not possible to measure the power losses in the cable separately, thus these values must be taken only as a reference. This way, Figure 13 shows that the results provided by S3D FEM models are very close to the estimates derived from experimental measurements, hence reasserting the validity of proposed short model. Moreover, these results also verify that the expressions of the IEC standard overestimate both loss factors when compared to experimental estimates and S3D simulations, especially $\lambda_2$. Additionally, it is also observed that 2D and 2.5D models do underestimate the screen's loss factor, especially in cables with magnetic armor. On the contrary, while 2D models overestimates the armor's loss factor, 2.5D models do the opposite, resulting in almost neglectable armor losses. In any case, from Figure 13 it can be concluded that, as expected, armor losses are quite low, far from the IEC estimations, but not negligible as provided by 2.5D FEM models.

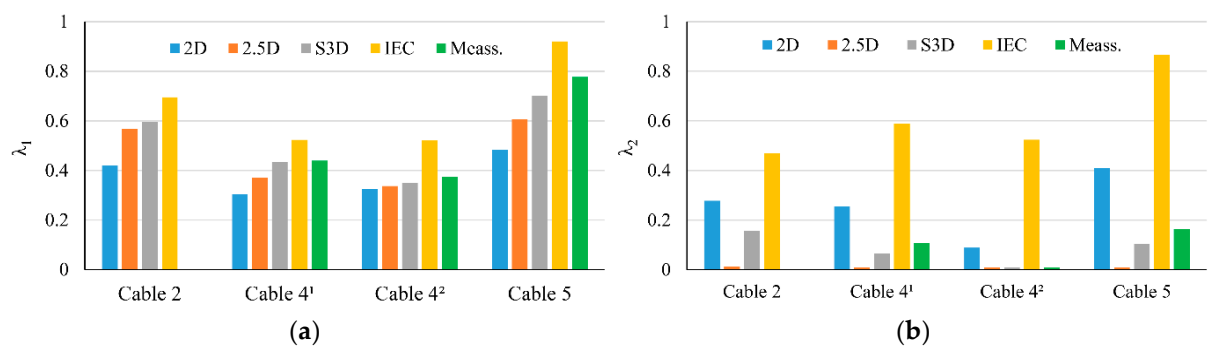

**Figure 13.** Loss factor obtained from FEM models, IEC standard and experimental measurements (solid bonding) for: (**a**) the screens; (**b**) the armor.

## 5. Conclusions

This work evaluates different ways for the simulation of three-core armored cables by means of 3D FEM models. To this aim, up to five MV and HV cables are simulated by means of full periodic and non-periodic 3D FEM models. This way, the accuracy of non-periodic models is analyzed, and the minimum model length for having accurate results is derived. From this analysis it concluded that a model length of about 125 % of the *CP* of the cable should be employed for having errors below 10 % in sheath and armor losses.

On the other hand, this work also analyzes the adequate length and boundary conditions for obtaining the shortest model possible with the same accuracy of full periodic models. As a result, a new way to simulate three-core armored cables is proposed, where rotated periodic boundary conditions are to be applied at both ends of a model as long as the *CP* of the cable. The accuracy of this new proposal is compared to full periodic models, resulting in errors below 0.5 % in *R*, *X*, $I_s$, $\lambda_1$, and $\lambda_2$. Furthermore, a great reduction in the simulation time is obtained in relation to full periodic models (more than 90 %) and non-periodic models (about 15 %).

Finally, the results in *R* and *X* derived from the new method in 145 kV and 245 kV cables are compared to experimental measurements, resulting in a very good agreement (differences below 3 % in most of the cases). Additional comparisons are performed with the IEC standard and power losses estimates derived from the experimental measurements, concluding that the new proposal fits very well the experimental estimates and validating the well-known conclusion regarding the overestimation of the power losses derived from the IEC standard.

As a conclusion, the new method here proposed brings a new and more effective tool to provide very important data and valuable knowledge about all phenomena that take place inside three-core armored cables, very useful for saving time in cable design and the development of new accurate analytical expressions for the current rating of these power cables.

**Author Contributions:** J. C. P-L conceived the idea of the project, undertook the modelling work and wrote the paper, P. C-R. obtained funding and thoroughly reviewed the paper, and M. H. provided valuable data, gave helpful comments and revised the paper.

**Funding:** This research was funded by the Agencia Estatal de Investigación and Fondo Europeo de Desarrollo Regional (AEI/FEDER, UE) under the project ENE2017-89669-R and by the Universidad de Sevilla (VI PPIT- US) under grant 2018/00000740.

**Conflicts of Interest:** The authors declare no conflict of interest.